\documentclass[fleqn,usenatbib]{mnras}

\usepackage{newtxtext,newtxmath}

\usepackage[T1]{fontenc}
\usepackage{CJK}

\DeclareRobustCommand{\VAN}[3]{#2}
\let\VANthebibliography\thebibliography
\def\thebibliography{\DeclareRobustCommand{\VAN}[3]{##3}\VANthebibliography}

\usepackage{graphicx}	
\usepackage{amsmath}	
\usepackage[dvipsnames]{xcolor}

\title[Off-axis MeV/TeV photons from structured GRB jets]{Off-axis MeV and very-high-energy gamma-ray emissions from structured gamma-ray burst jets}

\author[\v Z. Bo\v snjak et al.]{
\v Zeljka Bo\v snjak,$^{1}$\thanks{E-mail: zeljka.bosnjak@fer.hr(ZB)}
B. Theodore Zhang (张兵)\thanks{E-mail: bing.zhang@yukawa.kyoto-u.ac.jp (BTZ)},$^{2}$
Kohta Murase\thanks{E-mail: murase@psu.edu (KM)},$^{3,4,5,6,2}$
and Kunihito Ioka\thanks{E-mail: kunihito.ioka@yukawa.kyoto-u.ac.jp (KI)}$^{2}$
\\
$^{1}$Faculty of Electrical Engineering and Computing, University of Zagreb, Unska ul. 3, 10000 Zagreb, Croatia\\
$^{2}${Center for Gravitational Physics and Quantum Information, Yukawa Institute for Theoretical Physics, Kyoto University, Kyoto, Kyoto 606-8502, Japan}\\
$^{3}${Department of Physics, The Pennsylvania State University, University Park, PA 16802, USA}\\
$^{4}${Department of Astronomy \& Astrophysics, The Pennsylvania State University, University Park, PA 16802, USA}\\
$^{5}${Center for Multimessenger Astrophysics, Institute for Gravitation and the Cosmos, The Pennsylvania State University, University Park, PA 16802, USA}\\
$^{6}${School of Natural Sciences, Institute for Advanced Study, Princeton, NJ 08540, USA}
}

\date{Accepted XXX. Received YYY; in original form ZZZ}

\pubyear{2023}

\begin{document}
\begin{CJK*}{UTF8}{gbsn}
\label{firstpage}
\pagerange{\pageref{firstpage}--\pageref{lastpage}}
\maketitle

\begin{abstract}
Very-high-energy (VHE) photons around TeV energies from a gamma-ray burst (GRB) jet will play an essential role in the multi-messenger era, with a fair fraction of the events being observed off-axis to the jet. We show that different energy photons (MeV and TeV photons in particular) arrive from different emission zones for off-axis observers even if the emission radius is the same. The location of the emission region depends on the jet structure of the surface brightness, and the structures are generally different at different energies, mainly due to the attenuation of VHE photons by electron-positron pair creation. This off-axis zone-shift effect does not justify the usual one-zone approximation and also produces a time-delay of VHE photons comparable to the GRB duration, which is crucial for future VHE observations, such as by the Cherenkov Telescope Array. 

\end{abstract}

\begin{keywords}
radiation mechanisms: general -- relativistic processes -- stars: jets -- transients: gamma-ray bursts
\end{keywords}



\section{Introduction}
The very-high-energy (VHE; $\gtrsim 0.1\rm~TeV$) emission has been observed by Imaging Atmospheric Cherenkov telescopes (IACTs) from several gamma-ray bursts  
(GRB 180720B, GRB 190114C, GRB 190829A, GRB 201015A, GRB 201216C and GRB 221009A; for a recent review see e.g., \cite{Noda:2022hbo}).  
Among these, GRB 190829A and GRB 201015A are classified as low-luminosity (LL) GRBs based on their prompt emission luminosity.
For short gamma-ray bursts (sGRBs), the MAGIC collaboration has reported a significance of $\sim 3\sigma$ detection of gamma-rays from sGRB 160821B~\citep{MAGIC:2020ikk}. 
The detection of GRBs at the VHE band provided new clues for the understanding of the physics of GRBs, as modeling the emission on a broad energy range from optical to TeV allows us to constrain the microphysics parameters in the emission regions~\citep[e.g.,][]{derishevpiran2021,asano2020}. 
Even though the origin of VHE gamma-rays from GRBs is consistent with the synchrotron self-Compton (SSC) afterglow model for current observations~\citep[e.g.,][]{Meszaros:1994sd, Zhang:2001az, sari01}, the detection of VHE gamma-rays is expected also during the GRB prompt emission phase~\citep[e.g.,][]{inoue2013, vurm2017, bosnjak09, banerjee22, gillgranot22}.  

The detection of gravitational wave (GW) source GW170817 associated with the short GRB 170817A \citep{abbott2017a, abbott2017b} opened a new era in the electromagnetic counterpart search to GRBs \citep{lambkobayashi17, iokanakamura2018, nakar2020}, as the associated electromagnetic signal potentially allows the identification of the host galaxy and the redshift measurement. 
The properties of sGRB 170817A were rather uncommon: its $\gamma$-ray luminosity ($\sim$ 10$^{47}$ erg/s) was four orders of magnitude lower than typical short GRB, and the prompt emission consisted of two distinct components showing the unexpected spectral evolution. 
The first pulse was fitted with a cutoff power-law model with a hard low-energy spectrum, and it was followed by the second pulse dominated by a thermal emission \citep{pozanenko2018}. The early observations in the optical/near-infrared band were interpreted as quasi-thermal radiation from a kilonova \citep{tanaka17, kasen2017, utsumi2017, pian2017}. 
The radio and X-ray observations exhibited a gradual rise in the emission.
Early-time X-ray and radio observations were consistent with a wide-angle, mildly relativistic ($\Gamma\sim$ 2-3) cocoon emission \citep{kasliwal2017}. The VLBI observations of the superluminal motion suggested that the late-time emission was dominated by the narrowly-collimated ($\theta_c \lesssim$ 5$^\circ$) jet,  observed from a large viewing angle $\sim$14$^\circ$-28$^\circ$ \citep{mooleydeller18, ghirlanda19}.  As a viable scenario for the interpretation of the steadily rising afterglow luminosity in radio to X-ray \citep{margutti2018, mooley2018}, the structured jet was discussed \citep{Ioka:2019jlj, lazzati2018, urrutia2021, takahashi2021}.
 
The angular structure of the GRB jets was proposed early on \cite[e.g.,][]{meszarosreeswijers98, zhanga, rossi03, kumargranot2003, granotkumar2003} as it may be arising  during the jet launch or during the interaction of the jet with the dense environment following the merger of a neutron star binary \citep{aloy2005, murguiaberthier2017, kathirgamaraju2018, preau2021, gottlieb2021}; for a recent review see \cite{salafiaghirlanda2022}. 

In this work, we study the emission zone of VHE gamma-rays for a structured jet, similar to GW170817/GRB 170817A when viewed off-axis. We focus on the model described in \cite{Ioka:2019jlj}, where the off-axis emission arrives largely from the off-center jet when the jet luminosity is decreasing sharply outward as it is required from the observations of GRB 170817A. \cite{matsumotoB} revisited the compactness of the gamma-ray sources given by \cite{lithwick01} for arbitrary viewing angles, and confirmed that the relativistic jet core cannot be the origin of the observed emission in GRB 170817A. Future VHE facilities such as CTA will allow the follow-up of the gravitational events in the VHE band, and GW170817-like objects are promising sources of off-axis VHE gamma rays \citep{murase+2018}. We apply these findings in the study of the observed surface brightness of the jet emission taking into account the opacity of the source to gamma-rays. 
Recently, \cite{hendriks22} simulated a population of binary neutron stars observed by GW detectors (LIGO, Virgo, the Einstein Telescope and the Cosmic Explorer) and made predictions for the detection of sGRBs by {\it Fermi}/GBM, {\it Swift}/BAT and GECAM using a top-hat jet model for a GRB. Our study can also be implemented in such simulations to make predictions for future VHE observations.

This kind of study can be interesting also for LL GRBs \citep{Soderberg06,Liang07}. For LL GRBs often the relativistic shock breakout model is discussed \citep{Campana06,nakarsari2012, nakar15}, where the energy deposition is done by a narrow jet in the low-mass extended material. The induced shock is much less relativistic than the jet, and after the breakout produces the low-luminosity soft gamma-rays which are not narrowly beamed. The off-axis jet model was proposed for the interpretation of GRB emission properties in several events \citep{iokanakamura2001, yamazaki03, waxman2004, sato21}. LL GRBs are promising targets for future VHE facilities due to their predicted high local rate \citep{wanderman2010} and consequently, GeV/TeV observations being less affected by the extragalactic background light (EBL) attenuation~\citep{murase+2008,rudolph:2022}. The asymmetric collapse of massive stars may also be the source of GW emission \citep{shibata21}, where the interesting candidates are nearby low-luminosity GRBs \citep{kobayashimeszaros2003, daigne07, nakar15}. 

This paper is organized as follows: 
In Sec.~\ref{sec:structured-jet}, we describe the details of the off-axis structured jet model.
In Sec.~\ref{sec:vhe}, we discuss the origin of VHE gamma-rays and the optical depth due to the two-photon pair annihilation 
based on a structured jet model.
Our main results are presented in Sec.~\ref{sec:emission-zone}, where we arrive at the conclusion that different energy photons arrive from different emission zones in general.
In Sec.~\ref{sec:time-delay}, we discuss the detectability of the time delay.
In Sec.~\ref{sec:dis}, we study other effects that may affect our results and discuss the implications of this work.
Finally, we give a summary in Sec.~\ref{sec:sum}.

\section{Structured jet model}\label{sec:structured-jet}

\begin{figure}
    \centering
\includegraphics[width = 0.48\textwidth]{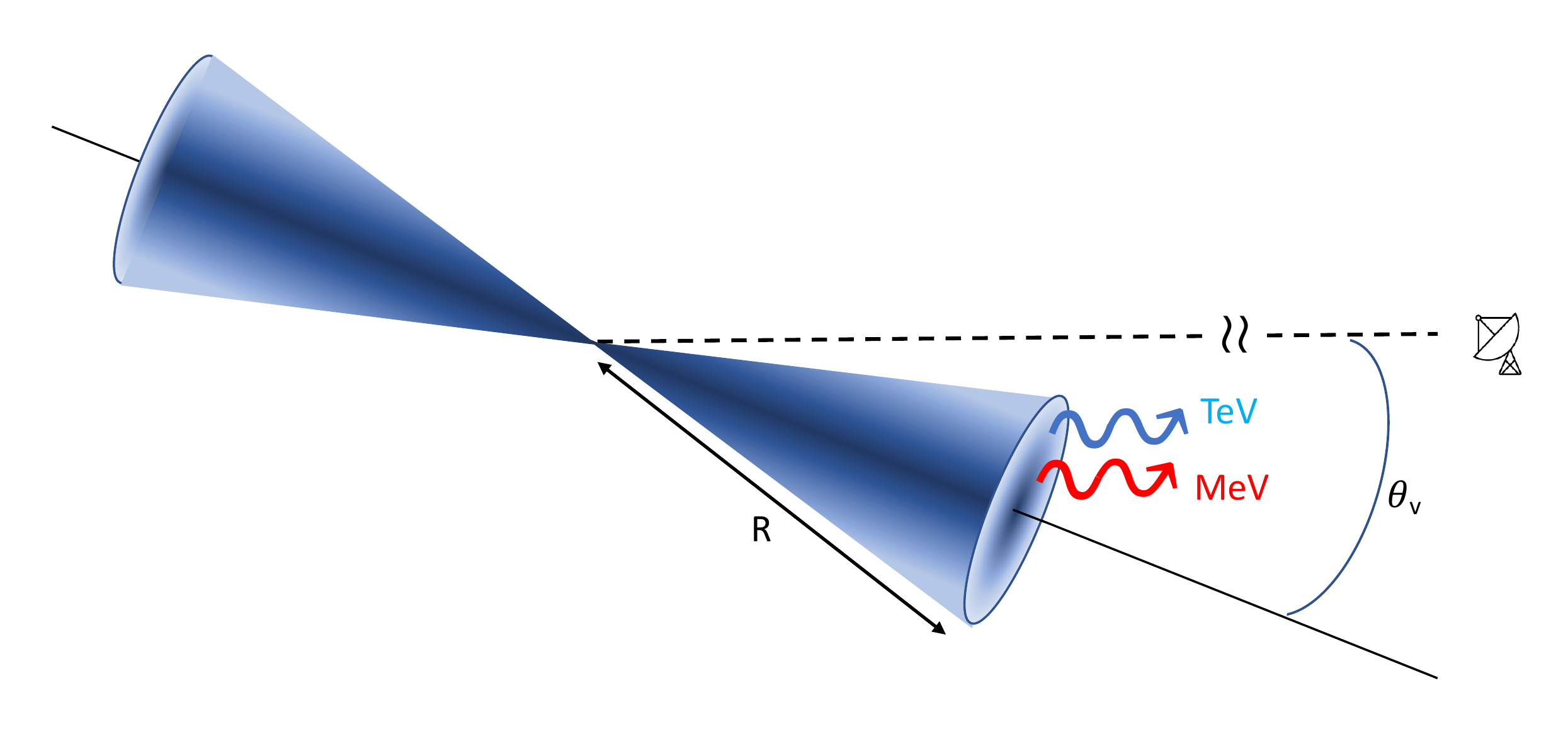}
    \caption{Prompt sub-MeV/MeV and VHE emission from a structured jet with viewing angle $\theta_v$. The distance of the emission region from the center is $R$ when measured in the laboratory frame. We show that different energy photons generally arrive from different emission zones for an off-axis observer.}
    \label{fig:scheme}
\end{figure}

In Fig.~\ref{fig:scheme}, we show a schematic picture of the emission from an off-axis structured jet that we applied: it consists of an energetic and highly relativistic core, with the energy and Lorentz factor sharply decreasing outwards.
Following~\cite{Ioka:2019jlj}, we consider an off-axis structured jet with a Gaussian shape,
\begin{equation}\label{eq:Eiso}
    E_\gamma(\theta) = \epsilon_\gamma E_0 \mathrm{exp}\left(-\frac{\theta^2}{2\theta_c^2}\right),
\end{equation}
where $E_\gamma(\theta)$ is the isotropic-equivalent radiation energy of the jet at an angle $\theta$ from the jet axis, $E_0$ is the isotropic equivalent energy measured along the jet axis, $\epsilon_\gamma$ is the radiation efficiency, and $\theta_c$ is the jet core opening angle (or the standard deviation of the Gaussian distribution).
The angle between the line of sight and a direction $(\theta, \phi)$ in the jet 
can be estimated as
\begin{equation}
    \mathrm{cos} \theta_\Delta = \mathrm{sin} \theta ~\mathrm{cos} \phi ~ \mathrm{sin} \theta_v + \mathrm{cos} \theta~\mathrm{cos} \theta_v,
\end{equation}
where $\theta_v$ is the viewing angle and $\phi$ is the azimuth angle with respect to the jet axis. In the following studies, we adopt fiducial values of $E_0 = 10^{52.8}\rm~erg$, $\theta_c = 0.059$ and $\theta_v = 0.38 \approx 22^\circ$ as inferred from the observations of sGRB 170817A~\citep{Troja:2018uns}.
We assume the radiation efficiency for sub-MeV/MeV prompt emissions is $\epsilon_\gamma = 0.1$ for simplicity. In reality, the radiation efficiency should have an angular structure $\epsilon_\gamma (\theta)$ which depends on the details of the radiative processes. In addition, we assume the Lorentz factor decreases outward with the shape described by the following relation,
\begin{equation}
    \Gamma(\theta) = \Gamma_{\rm max} \frac{1}{1 + (\theta/\theta_c)^\lambda},
\end{equation}
where $\Gamma_{\rm max} = 2000$ and $\lambda = 3.8$~\citep{Ioka:2019jlj}. 
We assume the energy of the structured jet dissipated at a distance $R$ from the explosion center in the laboratory frame, 
where the emission region could be described by a relativistic shocked shell with comoving width $\Delta^\prime \sim R / \Gamma (\theta)$. The corresponding timescales could be estimated as $\sim R / \Gamma (\theta)^2 c$.
However, the energy dissipation radius is difficult to predict without the knowledge of the jet composition and radial profile.
In this work, we consider the energy dissipation that occurred at the fixed radius $R$, and we will show that different energy photons generally arrive from different portions even for the same radius. 

\section{VHE prompt emission and optical depth}\label{sec:vhe}
In recent years, several GRBs have been observed at the VHE band, including both, high and low luminosity GRBs. All of the present VHE observations are consistent with the origin at the afterglow phase.
The unequivocal VHE emission during the prompt phase has not been detected yet, though it is possible that in the early VHE observations by MAGIC telescopes of GRB 190114C there was a contribution by the late prompt emission  \citep{nature1}. 
One of the difficulties faced when observing prompt emission is the duration of the response time following the alert system for IACTs.

In this work, we consider the detection of VHE emission during the prompt phase.
For the synchrotron emission from high-energy electrons, the maximum photon energy is $\sim 50\Gamma / (1 + z)\rm~MeV$ due to the limitation of the synchrotron energy loss process. 
The maximum photon energy could be somewhat increased considering the situation where high-energy electrons are accelerated far from the shock front where the magnetic field strength is lower but emit efficiently when these electrons travel near the shock front where the magnetic field strength is larger~\citep{Kumar:2012xm}. The inverse-Compton (IC) process would be more reliable to generate high-energy photons via upscattering low-energy photons to higher energies. The synchrotron self-Compton (SSC) process, where the same population of the non-thermal electrons produces synchrotron emission could also upscatter these photons to higher energies, and has been successful in explaining the observed VHE emission in the afterglow phase. 
However, the SSC process may not be efficient in the prompt phase due to the Klein-Nishina effect, where the IC cross-section for scattering decreases significantly. It occurs for the photon energies comparable to the electron rest mass when measured in the electron rest frame, $\gamma_e \nu_b^\prime \sim m_e c^2$, where $\nu_b^\prime$ is the target photon energy measured in the comoving frame. 
Nevertheless, if there are external photons, i.e., external thermal or non-thermal photons that overwhelm the prompt Band component in the low-energy range, the external inverse-Compton (EIC) process will dominate the high-energy photons at the VHE band. The origin of thermal photons could come from stellar emission or cocoon emission \citep[e.g.,][]{Toma09,decolle2018,Kimura19} and internal dissipation such as flares and extended emission \citep[e.g.,][]{murase+2018}.
The VHE gamma-rays in the prompt phase could also be contributed by the hadronic processes~\citep{Asano:2007my, Gupta:2007yb, razzaque2009, murase+2012, rudolph23}.

The low-energy (LE) sub-MeV/MeV prompt emission could be modeled with a spectral shape similar to the so-called Band function~\citep[e.g.,,][]{Ioka:2019jlj},
\begin{equation}
    f(\nu', \theta) = \frac{C}{\nu_0'(\theta)} \left(\frac{\nu'}{\nu_0'(\theta)}\right)^{1 + \alpha_B}\left[1 + \left(\frac{\nu'}{\nu_0'(\theta)}\right)^{2}\right]^{\frac{\beta_B - \alpha_B}{2}},
\label{eq:band}
\end{equation}
where the constant $C$ is chosen so that $\int d\nu^\prime f(\nu', \theta) = 1$.
We adopt the following relation for the observed peak energy $\nu_{0,\rm LE}(\theta, \phi)$ based on the observations of sGRB 170817A and the requirement to satisfy the Amati relation,  
\begin{equation}\label{eq:energy_comoving}
    \nu_{0,\rm LE}(\theta, \phi) = \delta_D \nu_{0,\rm LE}'(\theta) \approx 0.15 \times 10^3 \delta_D \left[1+\left(\frac{\theta}{\theta_c}\right)^{3.4}\right] {\rm~eV},
\end{equation}
where $\delta_D = 1 / (\Gamma (1 - \beta {\rm cos}\theta_\Delta))$ is the Doppler factor, and  $\nu_{0,\rm LE}'(\theta)$ is the peak energy in the comoving frame.

To model the energy spectrum at the VHE band of the prompt emission, we adopt a toy model which shares similar spectral properties as the low-energy prompt emission, with HE spectral peak
\begin{equation}\label{eq:energy_comoving}
     \nu_{0,\rm VHE}(\theta, \phi) = \delta_D \nu_{0,\rm HE}'(\theta) \approx 0.15 \times 10^9 \delta_D \left[1+\left(\frac{\theta}{\theta_c}\right)^{3.4}\right] {\rm~eV},
\end{equation}
where $\nu_{0,\rm HE}(\theta, \phi)$ is the observed peak energy of the VHE prompt emission. 

The optical depth for the two-photon pair annihilation process can be estimated as~\citep[e.g.,][]{Murase:2015xka},
\begin{align}\label{eq:tau}
    \tau_{\gamma\gamma} (\nu, \theta, \phi) &\approx n'(\tilde{\nu}') l^\prime \sigma_ {\gamma\gamma} \nonumber \\ &\approx  \frac{1} {4\pi R^2 \Delta R^\prime  \tilde{\nu}'} \frac{E_{\gamma} (\theta)}{\Gamma(\theta)} f(\tilde{\nu}', \theta)
  l^\prime \sigma_ {\gamma\gamma}
    \nonumber \\ &\simeq 1.6 \left(\frac{R}{10^{14}\rm~cm}\right)^{-2} \frac{\zeta(\theta)}{\zeta(4.5 \theta_c)},
\end{align}
where 
\begin{equation}\label{eq:zeta}
    \zeta(\theta) = (E_{\gamma} (\theta)/\Gamma(\theta))( f(\tilde{\nu}', \theta) / \tilde{\nu}').
\end{equation}
Also, $n'(\tilde{\nu}')$ is the comoving frame target photon number density at the characteristic energy $\tilde{\nu}' \approx \delta_D (m_e c^2)^2 / \nu$, $R$ is the distance of the emission region that is measured in the lab frame, $\Delta R^\prime \sim R /\Gamma(\theta)$ is the comoving shell width and the thickness of the gamma-ray emission region, $E_{\gamma} (\theta)/\Gamma(\theta)$ is the comoving radiation energy, $l^\prime$ is the photon path length before escaping from the source, and $\sigma_ {\gamma\gamma} \sim (7/12) \sigma_T$ is the cross section~\citep[e.g.,][]{1987MNRAS.227..403S}.
We simply assume $l^\prime \sim \Delta R^\prime$~\citep{matsumotoB}.
In the above estimates, we assume a broken power-law distribution with $\alpha_B = -1$ and $\beta_B = -2.5$.
We can see the optical depth depends on the distance of the emission region with $\tau_{\gamma \gamma} \propto R^{-2}$. 
The optical depth also depends on the polar angle, where $\zeta(\theta)$ sharply decreases with increasing $\theta$ when $\theta \gtrsim 2 \theta_c$.
In the above estimates, we assume $\theta = 4.5 \theta_c = 0.26 \approx 15^\circ$ and $\nu = 1\rm~TeV$. 
Note that we do not consider here the contribution to optical depth from scatterings of photons by e$^{\pm}$ created by the annihilation of photon pairs, as in case when this contribution was large we would expect the burst to be optically thick to all photons, independently on their energy \citep{lithwick01}.

The mean escape probability, or the attenuation factor, of the VHE emission, can be derived by solving the radiative transfer equation assuming uniform slab geometry~\citep{1987MNRAS.227..403S},
\begin{equation}
    \xi (\nu, \theta, \phi)= \frac{1 - e^{-\tau_{\gamma\gamma}(\nu, \theta, \phi)}}{\tau_{\gamma\gamma}(\nu, \theta, \phi)},
\end{equation}
which is $\xi \simeq 0.6$ for $\tau_{\gamma\gamma} = 1$ and $\xi \simeq 0.1$ for $\tau_{\gamma\gamma} = 10$.  In general, the optical depth $\tau_{\gamma\gamma}$ at a certain radius depends on the initial assumptions on Lorentz factor and energy profile, and on the Doppler factor.

\section{Different energy photons from different emission zones}\label{sec:emission-zone}
\begin{figure}
    \centering
\includegraphics[width = 0.5\textwidth]{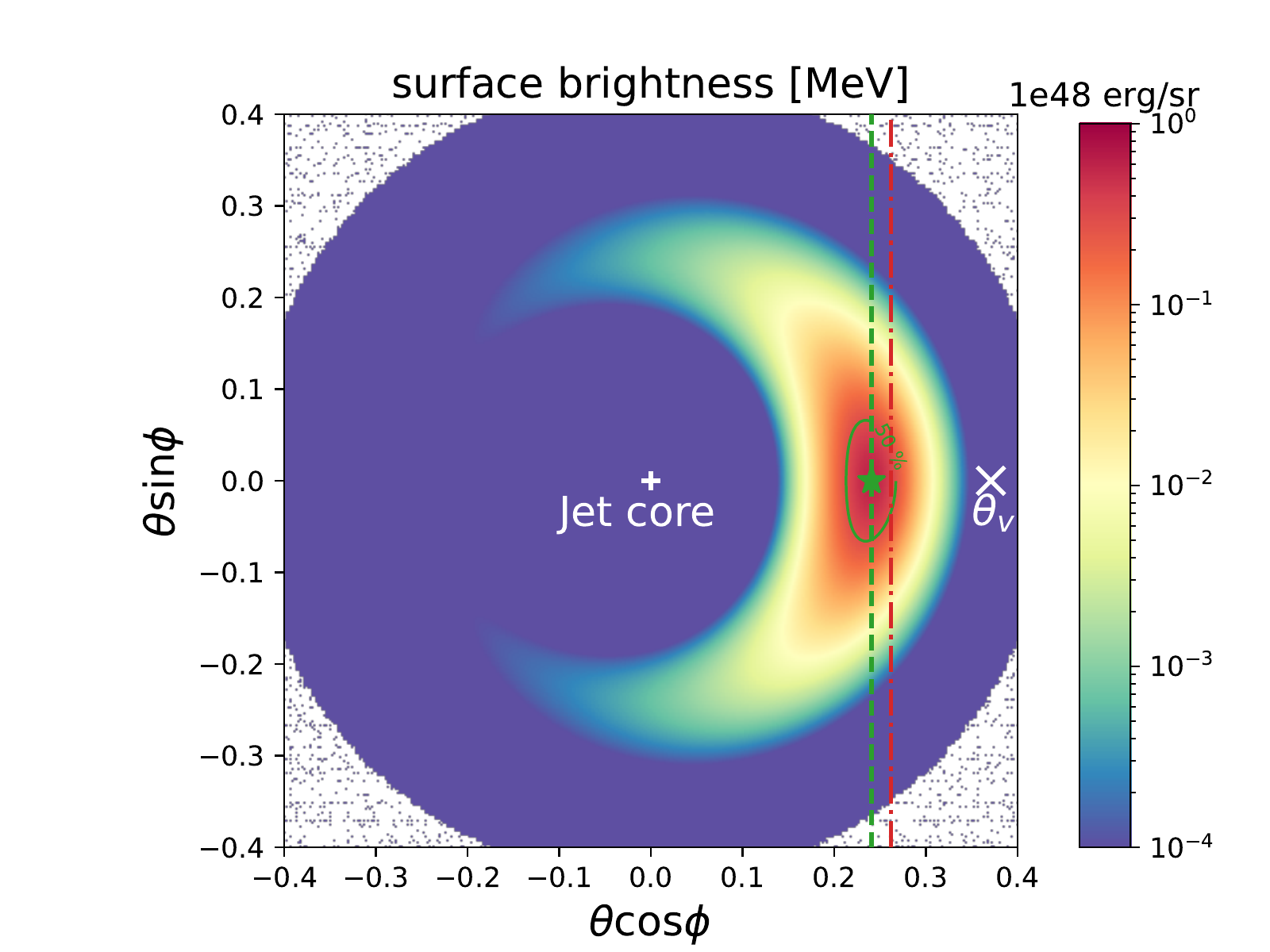}
\includegraphics[width = 0.5\textwidth]{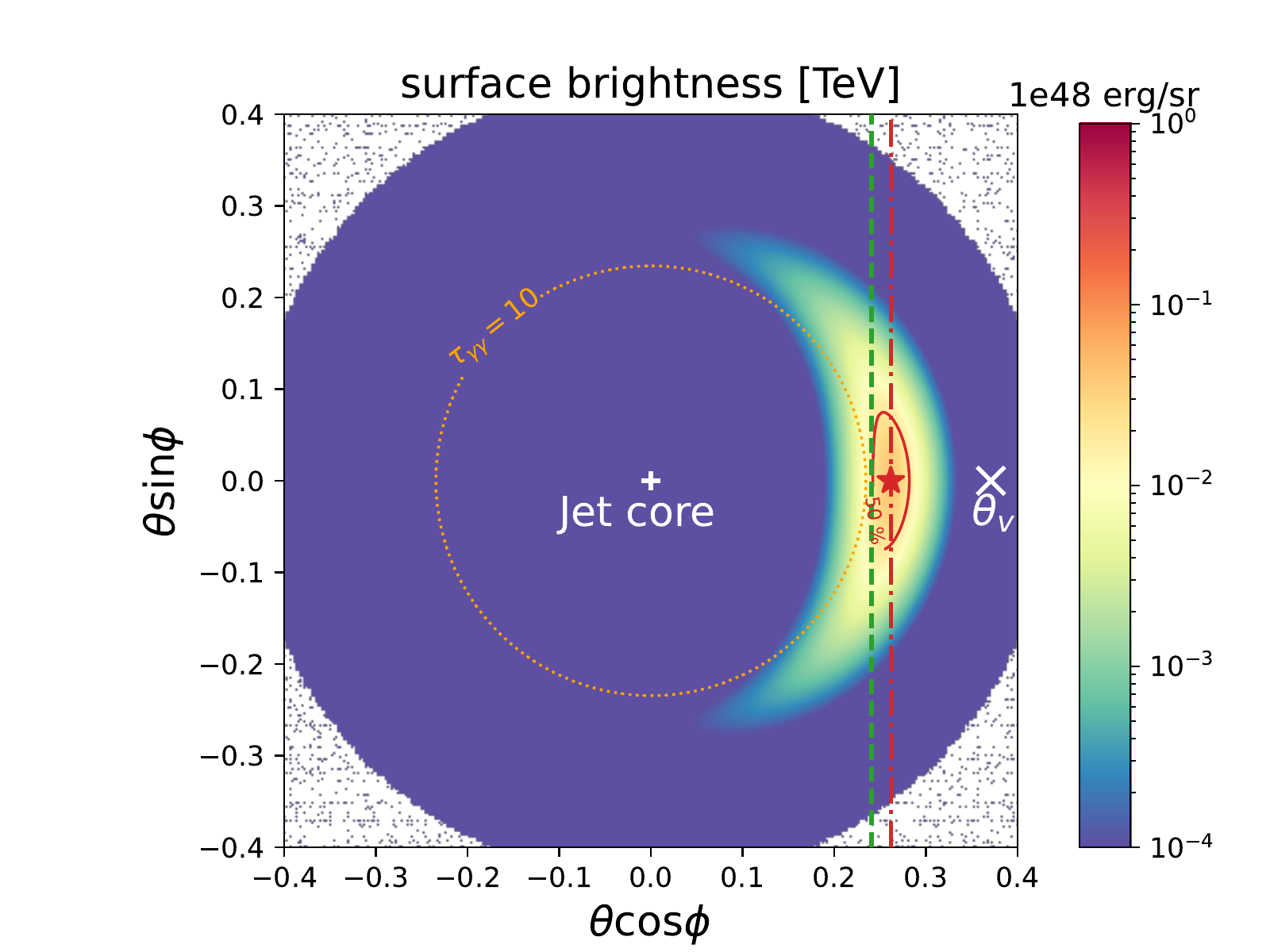} 
    \caption{The surface brightness distribution at the MeV band (upper panel) and TeV band (lower panel) of a structured jet for $\Gamma_{\rm max}$ = 2000 and $R = 10^{14}\rm~cm$. The position of the jet core is indicated as a white plus symbol and the viewing angle $\theta_v$ is marked as a white cross symbol. We can see that the emission regions with 50\% surface brightness are shifted between the MeV and TeV bands (see also Fig.~\ref{fig:surface_brightness_theta_phi}). This is mainly caused by the high optical depth of TeV gamma-rays ($\tau_{\gamma \gamma} = 10$ line with orange dotted line). The different emission region also leads to different arrival time in Fig.~\ref{fig:surface_brightness_delay}. 
    \label{fig:surface_brightness_g2000_r14}}
\end{figure}
\begin{figure}
    \centering
\includegraphics[width = 0.5\textwidth]{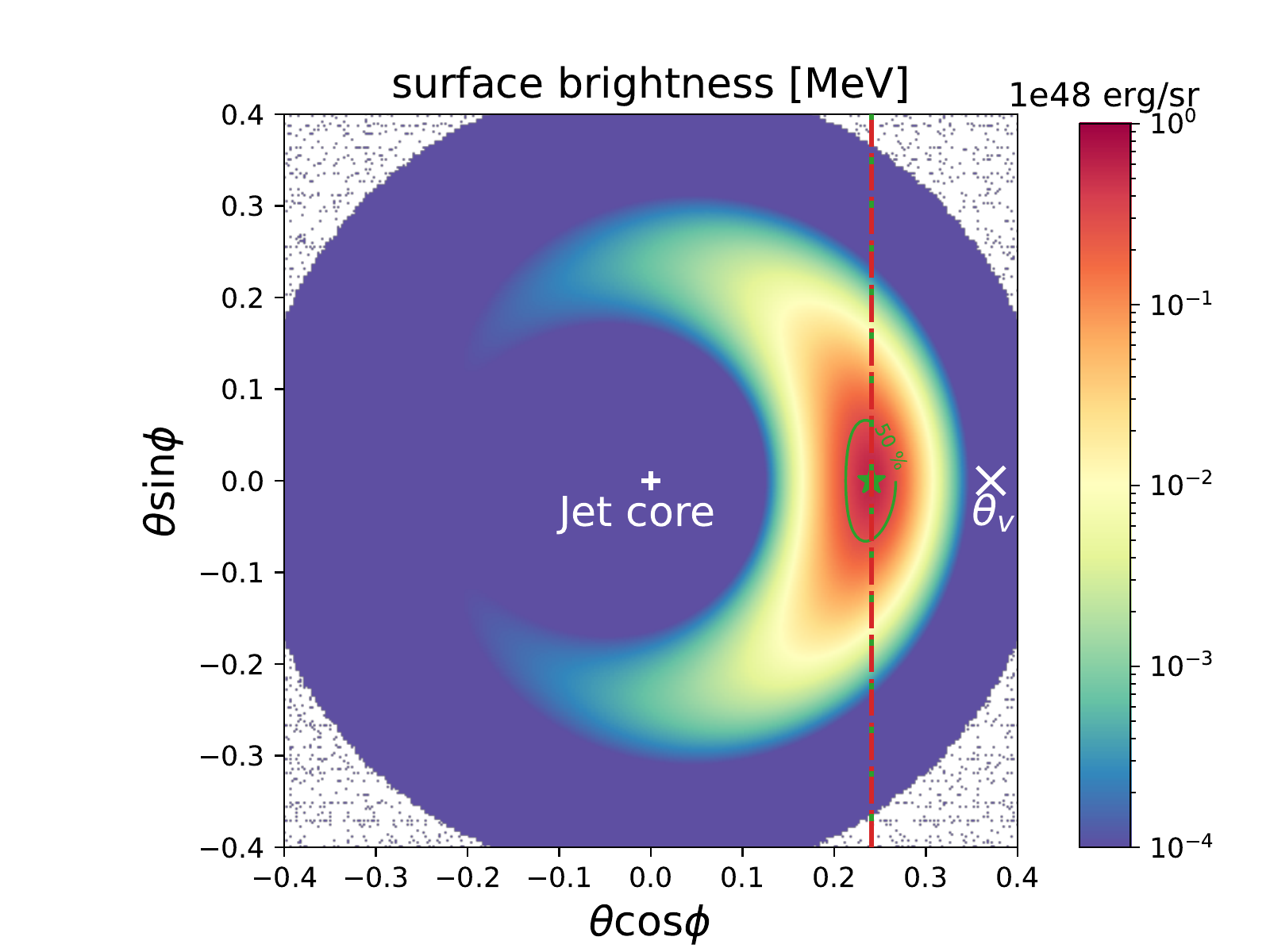}
\includegraphics[width = 0.5\textwidth]{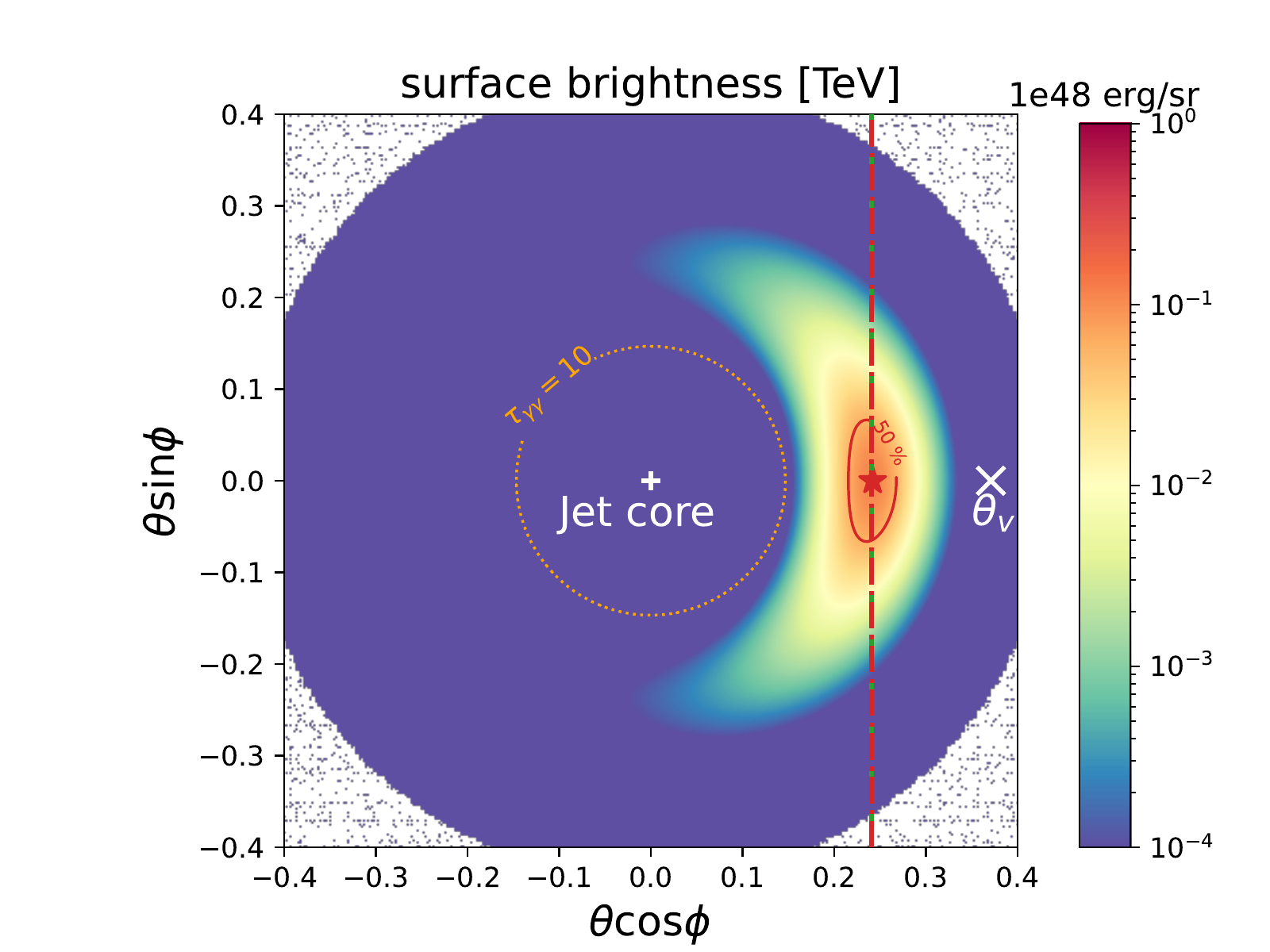} 
    \caption{Same as Fig.~\ref{fig:surface_brightness_g2000_r14}, but for $R = 10^{15}\rm~cm$. The shift of the peak position of the specific surface brightness between the MeV and TeV bands becomes much smaller.
    \label{fig:surface_brightness_g2000_r15}}
\end{figure}
The specific surface brightness per solid angle per frequency can be expressed as \citep{iokanakamura2018}
\begin{equation}\label{eq:surface_brightness}
    \frac{dE_{\gamma,\mathrm{iso}}}{d\Omega d\nu} = \frac{1}{4\pi}\frac{E_{\gamma}(\theta)[f(\nu, \theta, \phi; \nu_{0,\rm LE}) + f_{\rm HE}(\nu, \theta, \phi; \nu_{0,\rm HE})] \xi(\nu, \theta, \phi)}{\Gamma(\theta)^4[1-\beta(\theta) \mathrm{cos} \theta_\Delta]^3},
\end{equation}
where $\int d\nu f(\nu, \theta, \phi; \nu_{0,\rm LE}) = 1$ and $\int d\nu f_{\rm HE}(\nu, \theta, \phi; \nu_{0,\rm HE}) = 1/5$, and we assume the total energy radiated in the VHE energy band takes only $\sim 20\%$ of the total energy radiated in the low-energy sub-MeV/MeV band (this assumption would slightly increase the adopted value for radiative efficiency $\epsilon_\gamma$). Here $f_{\rm HE}(\nu, \theta, \phi; \nu_{0,\rm HE})$ has the same form as the low-energy spectrum given by Eq.~\ref{eq:band},  with the spectral peak in the comoving frame as in Eq.~\ref{eq:energy_comoving}. 

In Figs.~\ref{fig:surface_brightness_g2000_r14} and ~\ref{fig:surface_brightness_g2000_r15}, we show the surface brightness distribution, $\nu dE_{\gamma,\mathrm{iso}}/d\Omega d\nu$, on the jet surface at the MeV and TeV bands, respectively.
The peak position of the specific surface brightness is indicated by a green star at the MeV band (upper panel) and a red star at the TeV band (lower panel).
Due to the effect of the Doppler boost, the observed brightness for the off-axis structured jet is dominated by a small patch centered on the peak position with typical angular size $\Delta \Omega$, 
\begin{equation}
    \frac{dE_{\nu,\mathrm{iso}}}{d\nu} \approx \frac{dE_{\nu,\mathrm{iso}}}{d\Omega d\nu} \Delta \Omega,
\end{equation}
where $\Delta \Omega$ could be approximated as the region surrounded by the solid contours, 
which represents the position where the surface brightness decreases by a factor of $50\%$ compared to the peak value. The dotted circle in the lower panels is the position where the optical depth for TeV photons equals $\tau_{\gamma \gamma} = 10$. While the peak position of the surface brightness is shifted at the TeV band with respect to MeV band for smaller radius ($R$ = 10$^{14}$ cm, see Fig. \ref{fig:surface_brightness_g2000_r14}), this shift becomes much smaller at higher radii, see e.g. Fig. \ref{fig:surface_brightness_g2000_r15} for $R $ = 10$^{15}$ cm. However, the region where the surface brightness decreased by a factor of 50$\%$  becomes apparently larger, indicating the larger time spread of the arrival times for the TeV emission (this effect adds up to the difference in photon arrival times due to the different emission radii, see Fig. \ref{fig:surface_brightness_delay}).  Note that we did not include the evolution of different parameters (e.g. Lorentz factor, spectral properties) with radius in our calculation, while this may be expected with the jet propagation.

In conclusion, we could expect different energy photons to come from different emission zones for a structured jet when viewed off-axis. This is because the surface brightness distribution is different at different frequencies, mainly due to the different optical depth with more attenuation at the TeV band than the MeV band (and partly due to the different segment of the observed spectrum).

An important implication is that a popular one-zone approximation in the spectral analysis is not justified at all in the off-axis jet case.  
With the current facilities it is impossible to resolve the emission region for GRBs. One of the observable effects of such a phenomenon is the arrival time of photons from different emission regions, which we will discuss in the following section.

\section{Photon arrival time and possible time delay}\label{sec:time-delay}
For a relativistic structured jet, photons emitted at the same lab frame time $t$ at different locations may arrive at the observer at the same observed time T~\citep[e.g.,][]{2018pgrb.book.....Z}.
The observed time $T$ is related to the time in the laboratory frame $t$ as
\begin{equation}\label{eq:T-original}
    T = t - \frac{R}{c} {\rm cos}\theta_\Delta,
\end{equation}
where $R$ is the radius of the emitting shell measured in the laboratory frame.
The laboratory frame time $t$ can be estimated as
\begin{equation}\label{eq:lab-time}
    t = \int_0^R \frac{dr}{\beta(\theta) c} \approx \frac{R}{\beta(\theta) c}.
\end{equation}
The above approximation in Eq.~\ref{eq:lab-time} is valid if there is no acceleration or deceleration of the relativistic shell during the propagation.
The observed time $T$ is
\begin{align}\label{eq:obs-time}
    T &= \frac{R}{c \beta(\theta)} \left(1 - \beta(\theta) \mathrm{cos} \theta_\Delta\right) \nonumber \\
    &\approx 
    \begin{cases}
         \frac{R }{2 c \Gamma(\theta)^2} & (\theta_\Delta < 1/\Gamma(\theta)) \\
         \frac{R \theta_\Delta^2}{2 c} & (\theta_\Delta > 1/\Gamma(\theta))
    \end{cases},
\end{align}
where $\beta(\theta) \approx 1 - 1/2\Gamma^2$ and $\mathrm{cos}~ \theta_\Delta \approx 1- \theta_\Delta^2/2$.
From Eq.~\ref{eq:obs-time}, we can see that the observed time $T$ reaches a minimum value when $\theta_\Delta \sim 1/\Gamma$.

\begin{figure}
    \centering
\includegraphics[width = 0.5\textwidth]{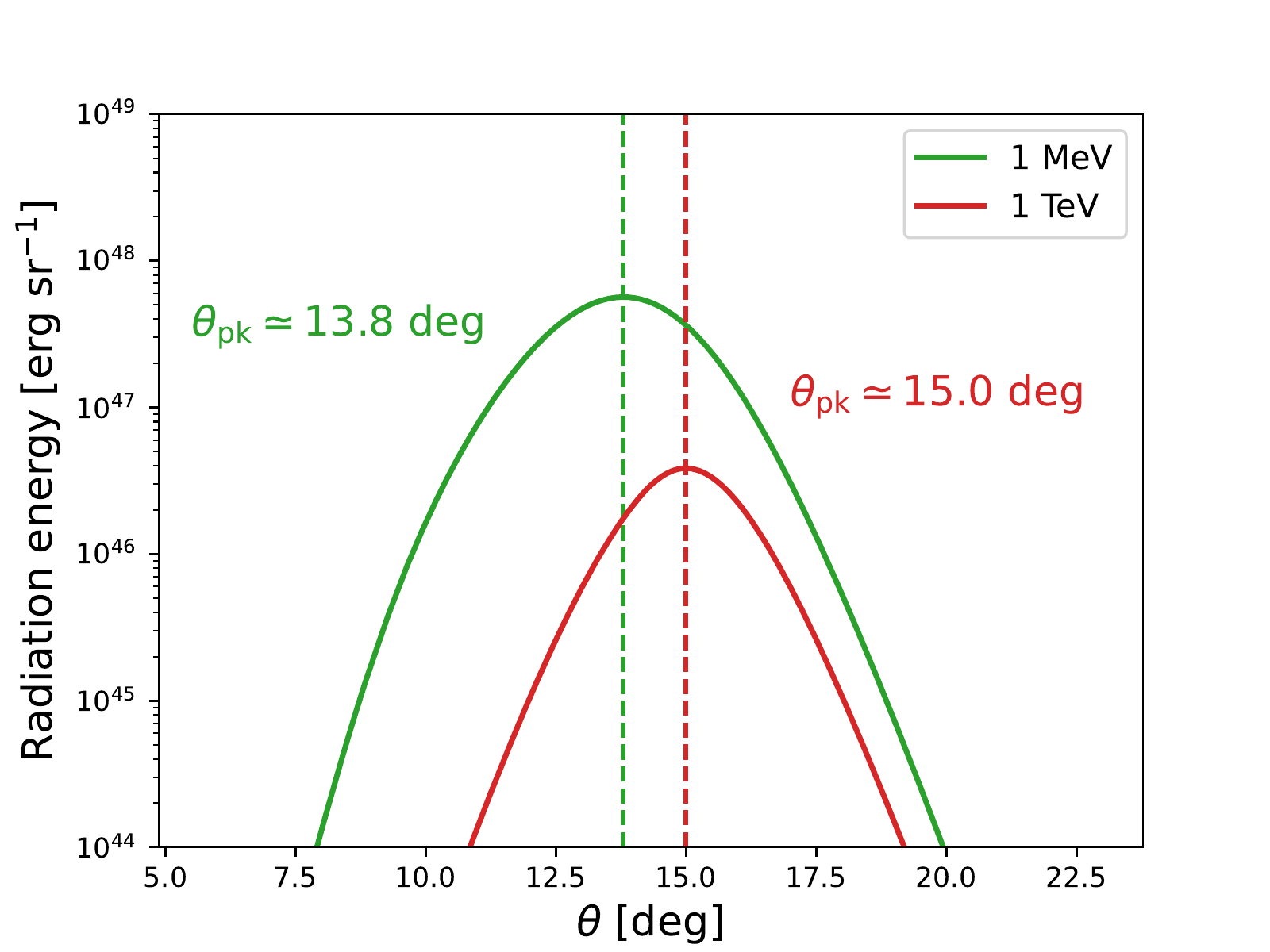}
    \caption{Similar to Fig.~\ref{fig:surface_brightness_g2000_r14}, we show the surface brightness distribution as a function of the polar angle $\theta$ when $\phi = 0$. 
    \label{fig:surface_brightness_theta_phi}}
\end{figure}

\begin{figure}
    \centering
\includegraphics[width = 0.5\textwidth]{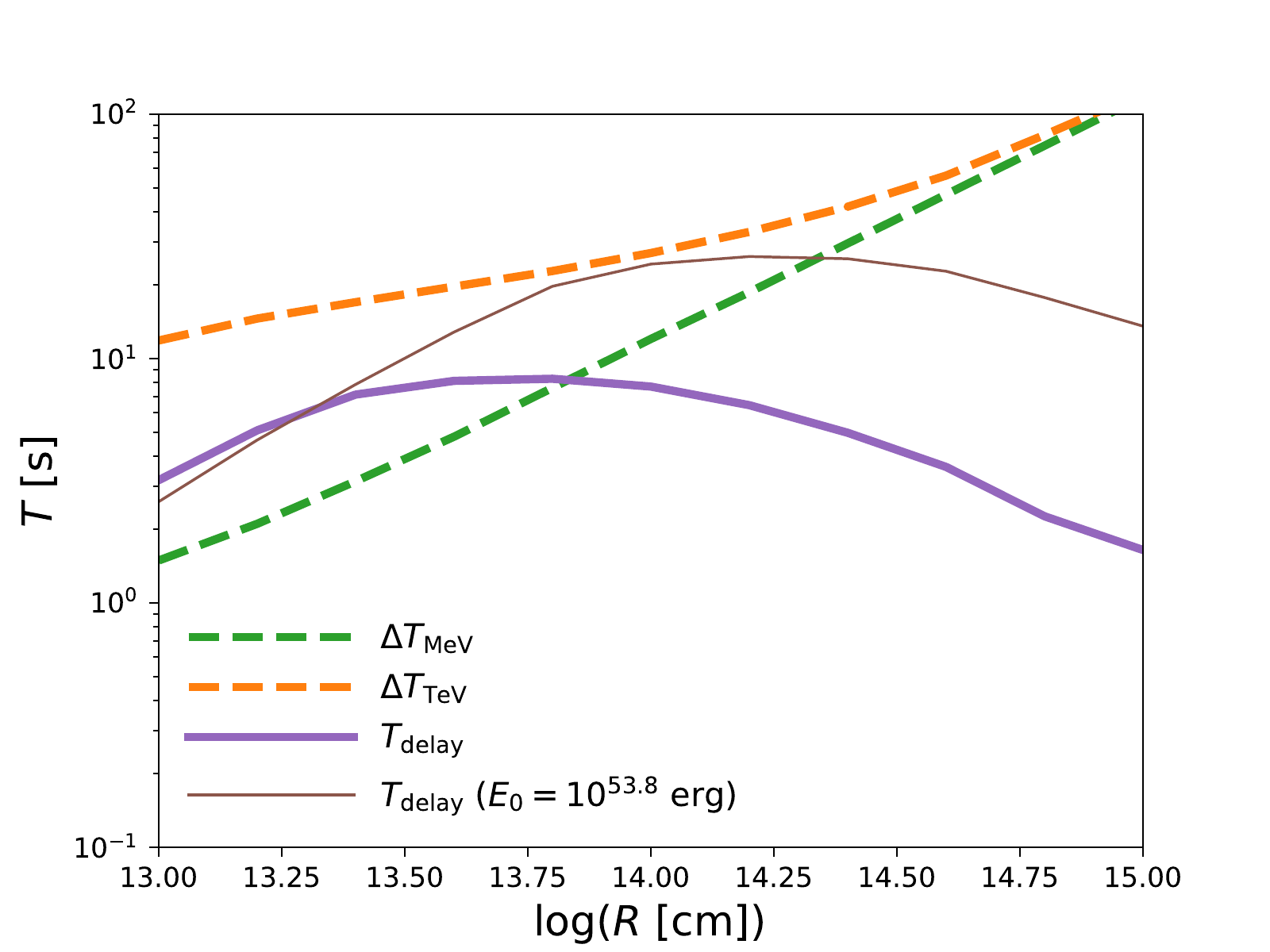}
    \caption{The arrival time delay between TeV and MeV photons and the duration of GRB prompt emission at the MeV energy band as a function of the emission region radius. The spectral and jet parameters are the same as in Fig.~\ref{fig:surface_brightness_g2000_r14}.
    Note that the viewing angle is set to $\theta_v=0.38$.
    \label{fig:surface_brightness_delay}}
\end{figure}

As shown in Fig.~\ref{fig:surface_brightness_theta_phi}, the peak position of the surface brightness along the direction of $\phi = 0$ is $\theta_{\rm MeV} \simeq 13.8\rm~deg$ at the MeV energy band and $\theta_{\rm TeV} \simeq 15.0\rm~deg$ at the TeV energy band, respectively.
The relative difference in the arrival time of TeV and MeV photons 
is shown in Fig.~\ref{fig:surface_brightness_delay} as a function of the emission radius.
Our results indicate that the arrival of the TeV photons is typically delayed compared to MeV photons, and the value reaches a maximum of approximately $T_{\rm pk, delay} \sim 8\rm~s$ when $R = 10^{13.75}\rm~cm$, see the purple thick solid curve in Fig.~\ref{fig:surface_brightness_delay}.

The time delay between the TeV photons and MeV photons is caused by the fact that the TeV photons and MeV photons have different surface brightness, mainly due to the two-photon annihilation optical depth, see Fig.~\ref{fig:surface_brightness_g2000_r14}.
The reason is that the emission zone of MeV photons is typically located at $\theta_{\rm MeV} \sim \theta_{\rm min}$, where $T$ reaches a minimum at $\theta_{\rm min}$. However, the emission zone of TeV photons is located at $\theta_{\rm TeV} > \theta_{\rm min}$, where $T$ is larger.
At a smaller radius, the whole VHE emission region is significantly attenuated due to the larger optical depth. 
With the increase of the emission radius, the VHE emission region near the center is still optical thick, but the outer region becomes transparent. Under such a situation, only VHE photons from the outer region could escape which are delayed compared to MeV photons.
The time delay between TeV and MeV photons becomes smaller for larger radius, i.e. $R \gtrsim 10^{15}\rm~cm$, where both the VHE emission region and sub-MeV/MeV emission region are optically thin.

We also show the time spread of the arrival times ( duration) of the MeV prompt emission 
from the interior of the half-maximum surface brightness line as a function of radius in dashed green line. Similarly, the duration of the TeV emission is shown in dashed orange line.
We can see that it is possible that the time delay between TeV and MeV photons could be larger than the typical duration of the MeV prompt emission.
The typical time delay between TeV and MeV photons could be significant for energetic events when $E_0 = 10^{53.8}\rm~erg$ because the VHE emission region move outwards due to the higher optical depth of TeV gamma-rays, see the brown thin solid curve.

With the current and near-future VHE gamma-ray facilities, especially with the operation of CTA~\citep{CTAConsortium:2017dvg}, one may expect co-incident detections of nearby GRBs at the VHE band with gravitational waves~\citep[e.g.,][]{murase+2018,Bartos:2019tsp}.
The study in this work suggests that the TeV emission pulse could lag behind the main pulse of the prompt emission at the MeV band, which could compensate for the large slewing time of CTA, which is $t_{\rm slew} \sim 20\rm~s$ for CTA-LST and $t_{\rm slew} \sim 90\rm~s$ for CTA-MST~\citep{banerjee22}. This effect, therefore, increases the probability of observing the TeV emission from short gamma-ray bursts during the prompt phase by CTA. Also, the possibility of the prolonged duration of the prompt phase may be interesting for interpreting the observed features of long GRBs linked to the compact object binary mergers, e.g. GRB 211211A \citep{troja22, jun22, mei22, rastinejad22}.

\section{Discussion and implications} \label{sec:dis}
When considering the off-axis structured jets, we found that different energy photons could arrive from different emission zones mainly due to the effect of the two-photon pair annihilation process. 
The main reason is that the optical depth for VHE photons is much higher in the core region on the jet surface, which gradually decreases outwards allowing VHE photons to escape.
In addition, we showed that the optical depth for VHE photons is sensitive to the emission radius, where the corresponding time delay between the typical arrival time of the TeV and MeV emission decreases with the increase of the emission radius.
Such a phenomenon could be prominent if the optical depth sharply decreases across the emission zone, such as in the case of the Gaussian jet adopted in this work, where $\zeta (\theta)$ strongly depends on $E_\gamma(\theta)$.
A similar effect is possible for the power-law structure of the jet energy if the power-law index is steep. 
Note that the angular dependence of the optical depth on the Lorentz factor $\Gamma(\theta)$ and energy spectrum $\nu_{0,\rm HE}^\prime(\theta)$ could also significantly affect the surface brightness distribution for different energy photons. The delayed arrival of $\gtrsim$ TeV photons is also expected in the EIC model \citep[e.g.,][]{murase+2010,Kimura19,zhang23}, without considering the structured jet.

Another important factor is the viewing angle $\theta_v$. If $\theta_v$ is close to or smaller than the jet core, we cannot resolve different emission regions which are similar to the on-axis case. However, if $\theta_v$ is too large, the received flux would be lower than the detection threshold.
The gravitational-wave data could provide an independent measurement of the inclination angle between the direction of the line-of-sight and jet axis~\citep{2020ApJ...893...38B}.
At present, we can only say that for an off-axis structured jet with properties similar to sGRB 170817A, different energy photons could originate from different emission zones.  \cite{banerjee22} estimated the minimum isotropic energy $\sim$ 10$^{47}$ ergs required for the detection of an event at z $\sim$ 0.1 (up to which the current GW detectors are expected to observe BNS mergers), and that would allow for this effect to be observed by the CTA.

\section{Summary}\label{sec:sum}
The off-axis model for the short gamma-ray burst GRB 170817A predicted that the most luminous region arises neither from the jet core around the primary axis, nor at the line of sight at the viewing angle $\theta_v$, but from the off-centre jet \citep{Ioka:2019jlj}. Adopting the same assumptions in this study, we showed that different energy photons could arrive to the observer from different emission zones for off-axis structured jets, and that the typical arrival time of VHE photons could be delayed compared to the typical arrival time of prompt sub-MeV/MeV photons. We discussed how the change in the emission radius could affect the VHE emission region and related arrival times.  Our results depend on the angular evolution of the total radiation energy $E_\gamma$ and of the Lorentz factor, and on the energy spectrum (currently the spectral evolution with radius was not accounted for). The off-axis structured jet could also be observed with a smaller or larger viewing angle depending on the energetics and detector threshold. One of the predictions of our model is the difference between the observed arrival time of prompt MeV emission and high-energy TeV emission. 
In general, the observation of the time arrival difference brings information on the emission radius. 

This model could be applicable to nearby short GRBs, VHE afterglow emission for energetic bursts, but also to LL GRBs, which are interesting as possible TeV emission and neutrino/UHECR sources \citep{murase+2008, mb10, boncioli2019, rudolph:2022}. 
As LL GRBs have presumably lower ejecta velocities and larger opening angles \citep{bromberg2011, cano2017,rudolph:2022}, the 
different assumptions should be taken into account for the estimates of the emission zones for MeV/TeV photons.

\end{CJK*}
\section*{Acknowledgements}
\v Z.B. acknowledges the support by FY2022 JSPS Invitational Fellowship for Research in Japan (Short-term) S22013. This work is partly supported by KAKENHI No. 23H05430, 23H04900, 22H00130, 20h01901, 20H01904, 20H00158 (K.I.).
The work of K.M. is supported by the NSF Grant No.~AST-1908689, No.~AST-2108466 and No.~AST-2108467, and KAKENHI No.~20H01901 and No.~20H05852.

\section*{Data Availability}
The data developed for the calculation in this work is available upon request.



\bibliographystyle{mnras}
\bibliography{example} 




\appendix


\bsp	
\label{lastpage}
\end{document}